\pdfoutput=1
\documentclass{PoS}

\usepackage{amsfonts,amsmath,graphicx,bm,slashed,color,mathrsfs,placeins,float}
\usepackage[utf8]{inputenc}
\usepackage[table]{xcolor}
\usepackage[caption = false]{subfig}

\definecolor{darkgreen}{rgb}{0.0,0.3,0.0}

\DeclareMathOperator{\arccot}{arccot}

\definecolor{Framableu}{RGB}{12,91,122}
\definecolor{Framableulight}{RGB}{18,144,176}
\definecolor{Framavert}{RGB}{142,156,72}
\definecolor{Framavertlight}{RGB}{227,235,199}
\definecolor{Framarouge}{RGB}{204,45,24}
\definecolor{Framarougelight}{RGB}{249,189,187}
\definecolor{Framaviolet}{RGB}{106,86,135}
\definecolor{Framavioletlight}{RGB}{211,197,232}
\definecolor{Framaorange}{RGB}{235,114,57}
\definecolor{Framaorangelight}{RGB}{235,209,197}
\definecolor{Framajaune}{RGB}{196,168,27}
\definecolor{Framajaunelight}{RGB}{255,235,181}
\definecolor{Framamarron}{RGB}{161,136,127}
\definecolor{Framamarronlight}{RGB}{215,204,200}
\definecolor{Framagris}{RGB}{97,97,97}
\definecolor{Framagrislight}{RGB}{245,245,245}

\newlength{\overwritelength}
\newlength{\minimumoverwritelength}
\newcommand{\overwrite}[3][red]{%
  \settowidth{\overwritelength}{$#2$}%
  \ifdim\overwritelength<\minimumoverwritelength%
    \setlength{\overwritelength}{\minimumoverwritelength}\fi%
  \stackrel
    {%
      \begin{minipage}{\overwritelength}%
        \color{#1}\centering\small #3\\%
        \rule{1pt}{9pt}%
      \end{minipage}}
    {\colorbox{#1!50}{\color{black}$\displaystyle#2$}}}

\newlength{\underwritelength}
\newlength{\minimumunderwritelength}
\newcommand{\underwrite}[3][red]{%
  \settowidth{\underwritelength}{$#2$}%
  \ifdim\underwritelength<\minimumunderwritelength%
    \setlength{\underwritelength}{\minimumunderwritelength}\fi%
  \stackrel
    {%
    {\colorbox{#1!50}{\color{black}$\displaystyle#2$}}
     \begin{minipage}{\underwritelength}%
        \rule{1pt}{9pt}\\%
        \color{#1}\centering\small #3%
     \end{minipage}}

}

\newcommand{\ltwopi}{\frac{L}{2\pi}}

\title{$K \pi$ scattering and the $K^*(892)$ resonance in 2+1 flavor QCD }

\ShortTitle{$K \pi$ scattering and the $K^*(892)$ resonance in 2+1 flavor QCD }

\author{\speaker{Gumaro Rendon}$^{a}$,
                Luka Leskovec$^{b}$, 
                Stefan Meinel$^{a,d}$, 
                John W. Negele$^{e}$,                 
                Srijit Paul$^{c,f}$, 
                Marcus Petschlies$^{g}$, 
                Andrew Pochinsky$^{e}$,
                Giorgio Silvi$^{i}$, 
                Sergey Syritsyn$^{d,h}$\\
      ${^a}$Department of Physics, University of Arizona, Tucson, AZ 85721, USA\\
      ${^b}$Theory Center, Jefferson Lab, Newport News, VA 23606, USA\\      
      ${^c}$Computation-based Science and Technology Research Center, The Cyprus Institute, 2121 Nicosia, Cyprus\\
      ${^d}$RIKEN BNL Research Center, Brookhaven National Laboratory, Upton, NY 11973, USA\\
      ${^e}$Center for Theoretical Physics, Massachusetts Institute of Technology, Cambridge, MA 02139, USA\\
      ${^f}$Department of Mathematics and Natural Sciences, University of Wuppertal, D-42119 Wuppertal, Germany\\
      ${^g}$Helmholtz-Institut f\"ur Strahlen- und Kernphysik, University of Bonn, D-53115 Bonn, Germany\\
      ${^h}$Department of Physics and Astronomy, Stony Brook University, Stony Brook, NY 11794, USA\\
      ${^i}$Forschungszentrum J\"ulich GmbH, J\"ulich Supercomputing Centre, 52425 J\"ulich, Germany\\
      E-mail: \email{jgrs@email.arizona.edu}}

\abstract{In this project, we will compute the form factors relevant for $B \to K^*(\to K \pi)\ell^+\ell^-$ decays. To map the finite-volume matrix elements computed on the lattice to the infinite-volume $B \to K \pi$ matrix elements, the $K \pi$ scattering amplitude needs to be determined using L\"uscher's method. Here we present preliminary results from our calculations with $2+1$ flavors of dynamical clover fermions. We extract the $P$-wave scattering phase shifts and determine the $K^*$ resonance mass and the $K^* K \pi$ coupling for two different ensembles with pion masses of $317(2)$ and $178(2)$ MeV.
}

\FullConference{The 36th Annual International Symposium on Lattice Field Theory - LATTICE2018\\
    22-28 July, 2018\\
    Michigan State University, East Lansing, Michigan, USA.}

\begin{document}

\FloatBarrier
\section{Introduction}
\FloatBarrier

\label{sec:introduction}

Processes of the type $b \to s \ell^+\ell^-$ occur at one-loop order and higher in the Standard Model and hence are suppressed. It is for this reason that these flavor-changing neutral-current decays are very important in the search for physics beyond the Standard Model.
The effective Hamiltonian that describes $b \to s \ell^+\ell^-$ decays at low energies \cite{Grinstein:1988me} has the operators
\begin{equation}
    O_{7(7')}  = \frac{m_b}{e}\bar{s} \sigma^{\mu\nu} P_{R(L)} b \: F_{\mu\nu}, \hspace{1ex}  O_{9(9')}  = \bar{s} \gamma_\mu P_{L(R)} b\: \bar{\ell} \gamma^\mu \ell, \hspace{2ex}    O_{10(10')}  = \bar{s} \gamma_\mu P_{L(R)} b\:\bar{\ell} \gamma^\mu \gamma_5 \ell,
\end{equation}
as well as four-quark and gluonic operators. The coefficients $C_i$ corresponding to these operators encode the short-distance physics and can be computed perturbatively in the Standard Model or in various new physics models. Global fits of experimental data from mesonic $b \to s \ell^+\ell^-$ transitions show deviations from Standard-Model predictions, in particular in the Wilson coefficient $C_9^{(\mu)}$ for the muonic final states \cite{Altmannshofer:2017fio,Capdevila:2017bsm}. One process that contributes significantly to this tension is $B \to K^{\ast}(\to K \pi)\ell^+\ell^-$. The hadron $K^{\ast}(892)$ in this process is unstable under the strong interaction, with a decay width of approximately $50$ MeV. The $B \to K^*$ form factors have been previously studied on the lattice in the single-hadron approach \cite{Horgan:2013pva,Horgan:2013hoa}, in which only a quark-antiquark interpolating field is used for the $K^*$ and the analysis is performed as if the $K^*$ were stable. This approach
has uncontrolled systematic uncertainties, which are expected to become more severe as the quark masses are lowered toward their physical values and the $K^*$ becomes broader.

To eliminate these systematic uncertainties, in this project we will compute the $B \to K \pi$ matrix elements of the relevant $b\to s$ currents, using the Brice\~no-Hansen-Walker-Loud generalization \cite{Briceno:2014uqa} of the Lellouch-L\"uscher formalism \cite{Lellouch:2000pv} to map the finite-volume matrix elements computed on the lattice to the desired infinite-volume $B \to K \pi$ matrix elements. The formalism requires a determination of the $K\pi$ scattering amplitude on the same lattice using L\"uscher's method \cite{Luscher:1990ux,Luscher:1991cf}. In this contribution, we present our preliminary results for this scattering amplitude. Previous lattice studies of $K\pi$ scattering can be found in Refs.~\cite{Beane:2006gj,Nagata:2008wk,Prelovsek:2013ela,Wilson:2014cna,Brett:2018jqw,Bali:2015gji}.

\FloatBarrier
\section{Calculating the scattering amplitude using lattice QCD}
\FloatBarrier

\label{sec:latscat}

L\"uscher's method \cite{Luscher:1990ux,Luscher:1991cf} utilizes the energy shifts caused by the interactions of the two-hadron system in the finite
lattice volume to extract the infinite-volume scattering amplitude. The cubic box with periodic boundary conditions in the spatial dimensions leads to
a quantization of the momenta and a purely discrete energy spectrum. In addition, the cubic box breaks the $SO(3)$ rotational symmetry, reducing it to $O_h$, or
to the relevant Little Group $LG^{\vec{P}}$ at nonzero total momentum $\vec{P}$. 

The operators that we use to extract the energy spectrum with the quantum numbers of the $K^*$ must have $I=1/2$ and must be projected to
irreps of $LG^{\vec{P}}$ that contain the $P$ wave (see Table \ref{tab:all_irreps}). We use operators with quark-antiquark and $K$-$\pi$ structure,
\begin{align}
& O_{\bar{q}q,i}(\vec{p})= \sum_{\vec{x}} e^{i \vec{p}\cdot \vec{x}}\overline{s}(x) \gamma_i  u(x), \quad
O_{\bar{q}q,0i}(\vec{p})= \sum_{\vec{x}} e^{i\vec{p}\cdot \vec{x}}\overline{s}(x) \gamma_0 \gamma_i  u(x) , \nonumber  \\
& O_{K\pi}\left(\vec{p}_1,\vec{p}_2\right)=\sqrt{\frac{2}{3}}\pi^{+}(\vec{p}_1)K^0(\vec{p}_2)-\sqrt{\frac{1}{3}}\pi^{0}(\vec{p}_2)K^+(\vec{p}_1) , 
\end{align}
where, for example, $K^+(\vec{p}_1)=\sum_{\vec{x}} e^{i\vec{p}_1\cdot \vec{x}}\bar s(x) \gamma_5 u(x)$. We project these operators to the required irreps
using the following procedure, which is based on the characters $\chi(R)$:
\begin{align}
O_{K\pi,\:\Lambda}^{\vec{P}} &= \dfrac{{\rm dim} (\Lambda)}{{\rm order}(LG^{\vec{P}})} \sum\limits_{R\in LG^{\vec{P}},\:\vec{m} \in Z^3} \chi(R)\:O_{K\pi}\left(\vec{P}/2 + R(\vec{P}/2+\frac{2\pi}{L}\vec{m}), \:\vec{P}/2-R(\vec{P}/2+\frac{2\pi}{L}\vec{m})\,\right) ,  \nonumber \\
O_{\bar{q}q,\:\Lambda}^{\vec{P}} &= \dfrac{{\rm dim} (\Lambda)}{{\rm order}(LG^{\vec{P}})} \sum\limits_{R\in LG^{\vec{P}}} \chi(R)\:R\:O_{\bar{q}q}(\vec{P}\,).
\end{align}

\begin{table}
\centering
\begin{tabular}{ |c|c|c|c|c|  }
 \hline
 $\ltwopi\vec{P}$& Little Group $LG^{\vec{P}}$ & Irrep $\Lambda^{\vec{P},r}$& Spin content & Dimension\\
 \hline
$(0,0,0)$   & $O_h$       & $A_{1g}$     &   $J=0,4,...$ & 1\\
\rowcolor{Framavertlight}
$(0,0,0)$   & $O_h$       & $T_{1u}$     &   $J=1,3,...$ & 3\\
$(0,0,1)$   & $C_{4v}$    & $A_1$  &   $J=0,1,...$ & 1\\
\rowcolor{Framavertlight}
$(0,0,1)$   & $C_{4v}$    & $E$          &   $J=1,2,...$ & 2\\
$(0,1,1)$   & $C_{2v}$    & $A_1$  &   $J=0,1,...$ & 1\\
\rowcolor{Framavertlight}
$(0,1,1)$   & $C_{2v}$    & $B_3$        &   $J=1,2,...$ & 1\\
\rowcolor{Framavertlight}
$(0,1,1)$   & $C_{2v}$    & $B_2$        &   $J=1,2,...$ & 1\\
$(1,1,1)$   & $C_{3v}$    & $A_1$  &   $J=0,1,...$ & 1\\
\rowcolor{Framavertlight}
$(1,1,1)$   & $C_{3v}$    & $E$          &   $J=1,2,...$ & 2\\
 \hline
\end{tabular}
\caption{List of the total momenta $\vec{P}$ that we use, along with their corresponding little groups, irreducible representations, and spin content \cite{Leskovec:2012gb}. Since we are interested in $J=1$, we only consider the highlighted irreps. } \label{tab:all_irreps}
\end{table}

We then use these operators to construct a correlation matrix, $C_{ij}(t) = \langle O_i(t)O^{\dagger}_j(0)\rangle$, for each of the irreducible representations.
There exist several methods to determine the spectrum from these correlation matrices; we use the generalized eigenvalue problem (GEVP) \cite{Blossier:2009kd,Orginos:2015tha}, where we solve the equation
\begin{align}
\label{eq:ortho}
C_{ij}(t)u^n_j(t) = \lambda_n(t,t_0) C_{ij}(t_0)u_j^n(t)
\end{align}
for the eigenvalues $\lambda_n(t,t_0)$ and eigenvectors $u_j^n(t)$. For large $t$, the eigenvalues satisfy $\lambda_n(t,t_0) \propto e^{-E_n t}$, and we perform single-exponential fits to $\lambda_n(t,t_0)$ to extract $E_n$.
Once we obtain the finite-volume spectrum, we map it to the infinite-volume phase shifts using the L\"uscher
quantization condition \cite{Luscher:1990ux}
\begin{equation}
\text{det} \left[ e^{2i\delta}\left(M^{\vec{P}}-i\right)-\left(M^{\vec{P}}+i\right)\right] = 0. 
\label{eq:det}
\end{equation} 
Here, $\left[ e^{2i\delta}  \right]_{lm,l'm'}\equiv e^{2i\delta_{l}} \delta_{ll'}\delta_{mm'}$ corresponds to the scattering matrix, where
$\delta_l$ is the scattering phase shift for angular momentum $l$, and $M$ is a known matrix of finite-volume functions depending on the the total
momentum $\vec{P}$, the scattering momentum, and the box size. When neglecting $D$-wave and higher contributions, it takes the form

\begin{align}
\label{eq:Mmat}
{M}^{\vec{P}} = \bordermatrix{~ & 0\,0        & 1\,0                    & 1\,1                       & 1\,-\!1\cr
                            0\:0 & w^{\vec{P}}_{00} & i\sqrt{3}w^{\vec{P}}_{10} & i \sqrt{3} w^{\vec{P}}_{11} & i\sqrt{3} w^{\vec{P}}_{1-1} \cr
                            1\:0 & -i\sqrt{3}w^{\vec{P}}_{10} & w^{\vec{P}}_{00}+2w^{\vec{P}}_{20} & \sqrt{3} w^{\vec{P}}_{21} & \sqrt{3}w^{\vec{P}}_{2-1} \cr
                            1\:1 & i \sqrt{3}w^{\vec{P}}_{1-1} & -\sqrt{3}w^{\vec{P}}_{2-1} & w^{\vec{P}}_{00}-w^{\vec{P}}_{20} & -\sqrt{6} w^{\vec{P}}_{2-2} \cr
                            1\,-\!1 & i \sqrt{3} w^{\vec{P}}_{11} & -\sqrt{3}w^{\vec{P}}_{21} & -\sqrt{6}w^{\vec{P}}_{22} & w^{\vec{P}}_{00}-w^{\vec{P}}_{20} \cr},\ 
\end{align}
where
\begin{align}
w^{\vec{P}}_{lm}\equiv \frac{1}{\pi^{3/2} \sqrt{2l+1} ~\gamma~ q^{l+1}}~ Z^{\vec{P}}_{lm}(1; q^2),
\end{align}
with the relativistic Lorentz factor $\gamma$, the generalized zeta function $Z_{lm}$, and the dimensionless scattering momentum $q=\frac{L}{2\pi}k$. The scattering momentum $k$ of the $n$-th energy level is determined by solving
\begin{equation}
\sqrt{s^{\vec{P},\Lambda}_n}=\sqrt{m_{\pi}^2+(k^{\vec{P},\Lambda}_n)^2}+\sqrt{m_{K}^2+(k^{\vec{P},\Lambda}_n)^2},
\end{equation}
where the center-of-mass energy $s^{\vec{P},\Lambda}_n$ is related to the energy $E^{\vec{P},\Lambda}_n$ on the lattice via $
\sqrt{s^{\vec{P},\Lambda}_n}=\sqrt{(E^{\vec{P},\Lambda}_n)^2-(\vec{P})^2}$.
The matrix ${M}^{\vec{P}}$ simplifies to a block-diagonal form once the specific symmetries belonging to the system with total momentum $\vec{P}$ are taken into account. The quantization condition (\ref{eq:det}) then becomes a product of quantization conditions, each belonging to its own irreducible representation $\Lambda$. 

The $K\pi$ system has no parity symmetry at non-zero momentum \cite{Prelovsek:2013ela}; consequently, certain irreps will have contributions from $S$-wave and $P$-wave scattering states at the same time. This becomes a challenging technical problem, because the $S$-wave phase shift is non-negligible in the region of interest. In this first analysis, we therefore limit ourselves to the irreducible representations that do not contain $l=0$, as indicated in Table \ref{tab:all_irreps}.
\FloatBarrier
\section{Parameters of the lattice gauge-field ensembles}
\FloatBarrier
We use two ensembles with $N_f=2+1$ dynamical quark flavors, utilizing a clover-improved Wilson action and the tadpole-improved tree-level Symanzik gauge action. The gauge links in the fermion action are smeared using one level of stout smearing to avoid instabilities in the hybrid-Monte-Carlo evolution. The parameters of the lattice gauge ensembles are given in Table \ref{tab:latinfo}.
\begin{table}[H]
\begin{center}
\begin{tabular}{|l|l|c|c|c|c|c|}
\hline
 Label        & $N_L^3\times N_t$ & $a\:(\rm fm)$ & $L$ (fm)  & $m_\pi$ (MeV) & $m_K$ (MeV) & $N_{\text{meas}}$ \cr
\hline
 \texttt{C13} & $32^3\times 96$   & $0.11403(77)$ & $3.65(2)$ & $317(2)$ & $527(4)$   & 7060 \cr
\hline
 \texttt{D6} & $48^3\times 96$   & $0.08766(79)$ & $4.21(4)$ & $178(2)$ & $514(5)$  & 5248 \cr
\hline
\end{tabular}
\caption{Parameters of the two ensembles and numbers of measurements analyzed. }
\label{tab:latinfo}
\end{center}
\end{table}
\FloatBarrier
\section{Preliminary results}
\FloatBarrier
\label{sec:results}
%

\begin{figure} 
    \centering
    \subfloat{%
        \includegraphics[width=0.45\textwidth]{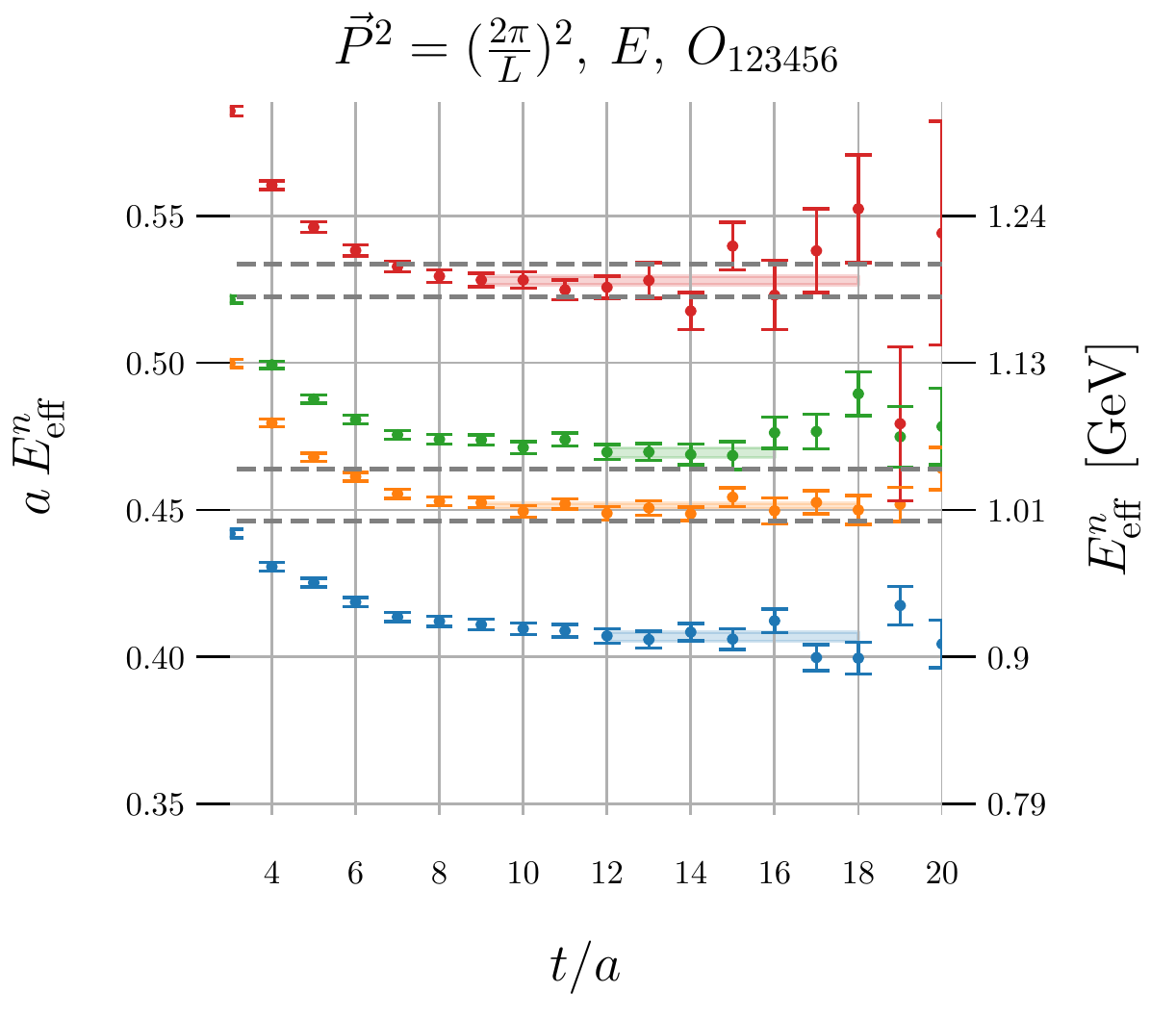}%
        }%
    \hfill%
    \subfloat{%
        \includegraphics[width=0.38\textwidth]{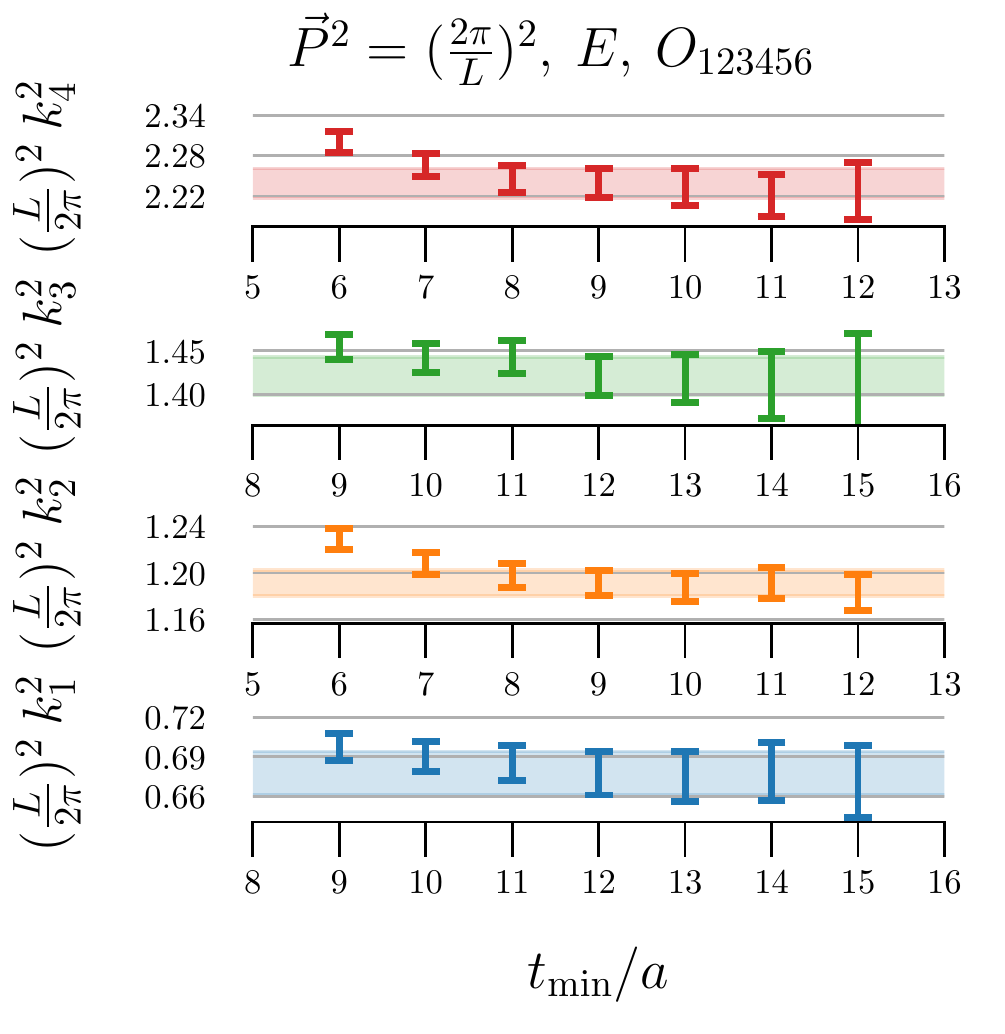}%
        }%
    \caption{Left: effective-mass plot of the GEVP eigenvalues for irrep $E$, which has a total momentum squared of $\vec{P}^2=\left(\frac{2\pi}{L}\right)^2$. Right: the squared dimensionless scattering momenta obtained from the fitted energies, plotted as a function of $t_{\rm min}$. The data shown here are from the \texttt{D6} ensemble.}
    \label{fig:meffsstability}
\end{figure}

Figure \ref{fig:meffsstability} (left) shows an example of an effective mass plot from the eigenvalues $\lambda_n(t,t_0)$ obtained with the GEVP method,
for the $E$ irrep with $\vec{P}^2=\left(\frac{2\pi}{L}\right)^2$. In Fig.~\ref{fig:meffsstability} (right) the stability of the single-exponential fits to
$\lambda_n(t,t_0)$ is demonstrated by varying the $t_{\rm min}$ at which the fits start. We choose the $t_{\rm min}$ at which the energies (or equivalently, the scattering momenta) reach stable plateaus.

\begin{figure}
\centering
  \includegraphics[width=0.65\linewidth]{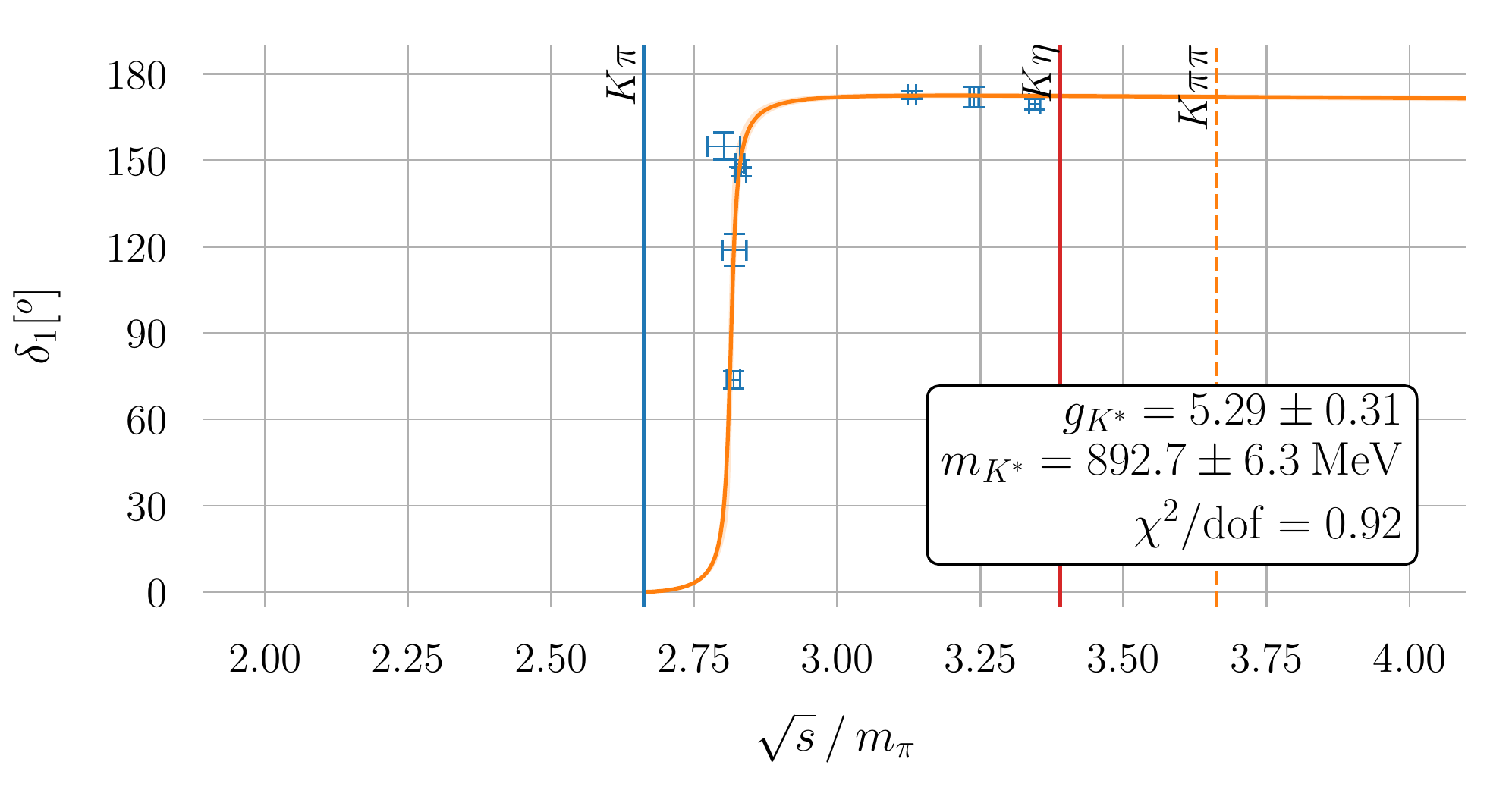}
  \caption{Phase shift points obtained from the \texttt{C13} ensemble, along with a Breit-Wigner fit using Eq.~(\ref{eq:BW}). The parameters obtained from the fit are given in the bottom-right corner. The relevant thresholds are indicated with vertical lines.}
  \label{fig:C13_ps}
\end{figure}

\begin{figure}
\centering
  \includegraphics[width=0.65\linewidth]{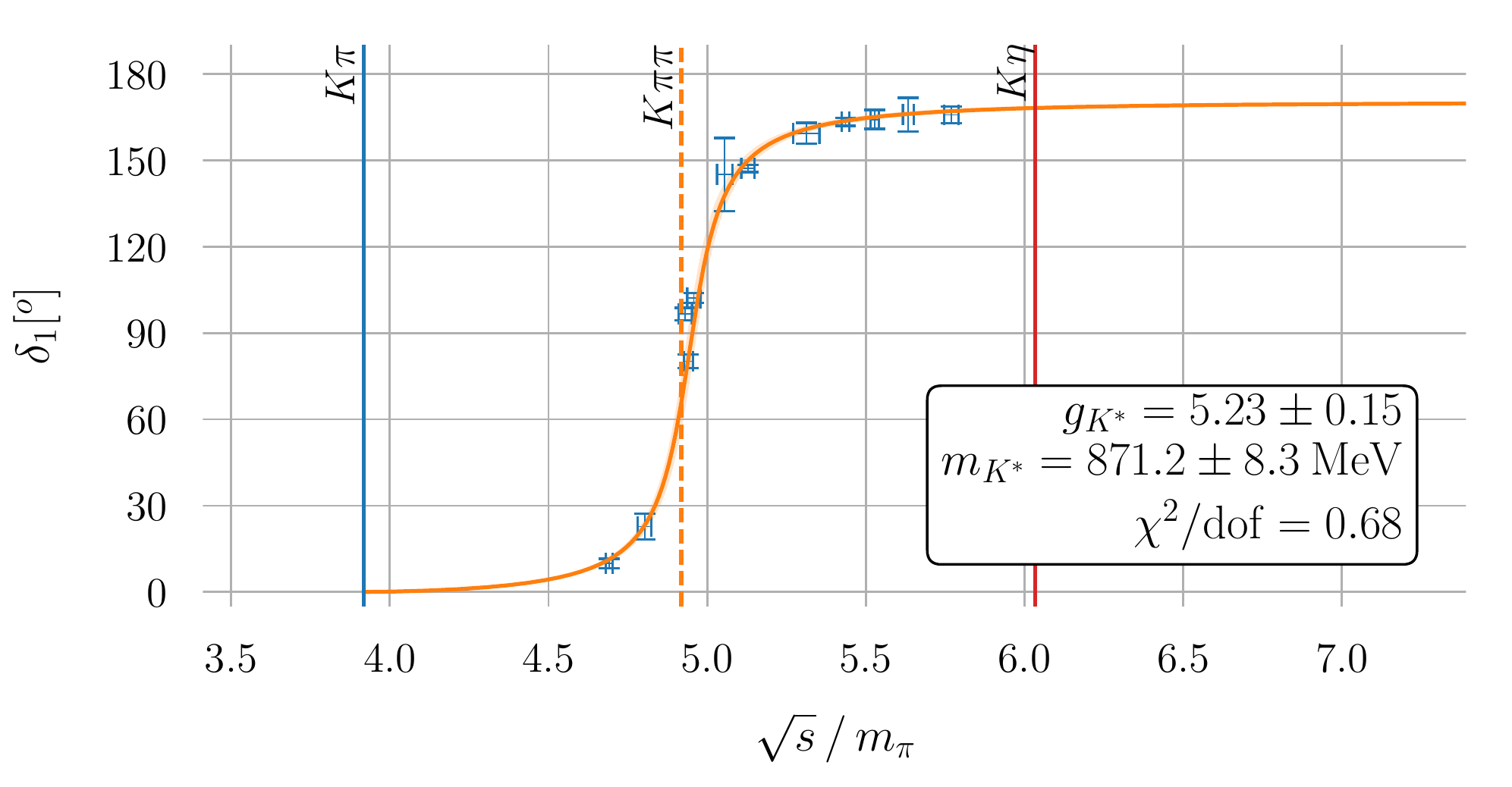}
  \caption{Like Fig.~\protect\ref{fig:C13_ps}, but for the \texttt{D6} ensemble.}
  \label{fig:D6_ps}
\end{figure}

The preliminary phase shift fit results obtained using L\"uscher's method are shown in Figs.~\ref{fig:C13_ps} and \ref{fig:D6_ps} for ensembles \texttt{C13} and \texttt{D6}, respectively. Also shown in the figures are fits of the energy-dependence using Breit-Wigner parametrizations of the form
\begin{equation}
\delta_{l=1}=\arccot{\left(\frac{6\pi\sqrt{s}}{g_{K^*}^2\, k^{3}}(m^2_{K^{*}}-s)\right)}.
\label{eq:BW}
\end{equation}
For both ensembles, we truncated the data before reaching the $K\eta$ threshold in order to avoid any inelastic effects. For the \texttt{D6} ensemble, the $K\pi\pi$ threshold is located near the resonance, as shown in Fig.~\ref{fig:D6_ps}. However, the coupling of this three-meson channel with the $K\pi$ channel is not observed in experiment \cite{Estabrooks:1977xe}, and it is therefore reasonable to assume that it does not significantly affect our elastic scattering results. The values of $g_{K^*}$ and $m_{K^*}$ obtained from our fits, also displayed in Figs.~\ref{fig:C13_ps} and \ref{fig:D6_ps}, are consistent with previous lattice calculations at similar pion masses \cite{Prelovsek:2013ela,Wilson:2014cna,Brett:2018jqw,Bali:2015gji}.


\section{Outlook}
\label{sec:outlook}

In the near future we plan to add a third ensemble of lattice gauge configurations, which will allow us to study the dependence on both $m_\pi$ and $a$.
Furthermore, we will include the irreducible representations in which the $S$ and $P$ waves mix, and we will include the coupling to the $K\eta$ channel.
After the determinations of the scattering amplitudes are completed, we will proceed with the analysis of the transition matrix elements for the process $B \to K \pi \ell^+\ell^-$, which is the main goal of this project.

\section{Acknowledgments}

We are grateful to Kostas Orginos and the other members of the William \& Mary and JLab lattice QCD groups for providing the lattice gauge ensembles. Computations for this work were carried out in part on facilities of the USQCD Collaboration, which are funded by the Office of Science of the U.S. Department of Energy, and on SuperMUC at Leibniz Supercomputing Centre, which is funded by the Gauss Centre for Supercomputing. SM and GR are supported by the U.S. Department of Energy (DoE), Office of Science, Office of High Energy Physics under Award Number DE-SC0009913. SM and SS also thank the RIKEN BNL Research Center for support. LL is supported by the U.S. Department of Energy, Office of Science, Office of Nuclear Physics under contract DE-AC05-06OR23177. JN was supported in part by the DOE Office of Nuclear Physics under grant DE-SC-0011090. AP was supported in part by the U.S. Department of Energy Office of Nuclear Physics under grant DE-FC02-06ER41444. SP is supported by the Horizon 2020 of the European Commission research and innovation programme under the Marie Sklodowska-Curie grant agreement No. 642069. We also acknowledge the use of the USQCD software QLUA for the calculations of the correlators.

\providecommand{\href}[2]{#2}\begingroup\raggedright\endgroup

\end{document}